\documentclass{sig-alternate-10pt}
\usepackage[utf8]{inputenc}
\usepackage{color,soul}
\usepackage{graphicx}
\usepackage{epstopdf}
\usepackage{url}
\usepackage{amsmath,amssymb,amsfonts}
\usepackage{caption,subcaption}
\usepackage{comment}

\usepackage[linesnumbered,ruled,vlined]{algorithm2e}
\usepackage[export]{adjustbox}
\usepackage{listings}
\usepackage{titlesec}
\usepackage{enumitem}

\graphicspath{{./figures/}}

\def\projectName{DistressNet-NG }
\def\ek{EdgeKeeper}

\titlespacing{\section}{3pt}{\parskip}{-\parskip}
\titlespacing{\subsection}{3pt}{\parskip}{-\parskip}
\titlespacing{\subsubsection}{3pt}{\parskip}{-\parskip}

\title{EdgeKeeper: Resilient and Lightweight Coordination for Mobile Edge Computing Systems}

\author{
S. Bhunia, R. Stoleru, M. Sagor, A. Haroon, A. Altaweel, M. Chao, M. Maurice$^{\dagger}$, R. Blalock$^{\dagger}$\\
\affaddr{Department of Computer Science and Engineering, Texas A\&M University}\\
\affaddr{$^{\dagger}$National Institute of Standards and Technology (NIST)}\\
\email{[sbhunia, stoleru, msagor, amran.haroon , altaweelala1983, chaomengyuan]@tamu.edu}
}

\begin{document}

\maketitle

\begin{abstract}
Mobile Edge Computing (MEC) has been gaining significant interest from first responders and tactical teams, primarily because they can employ handheld mobile devices to form a computing cluster (for computing tasks like face/scene recognition, virtual assistance) when connectivity to the cloud is not present or it is limited. High user mobility in first responder or tactical environments makes MEC challenging, as wireless links observe substantial fluctuations.  Typical cloud-based coordination (e.g., ZooKeeper-based service discovery and coordination, device naming, security) needed by edge computing tasks cannot work in these environments. Driven by the need for a resilient and lightweight coordination service, in this paper, we design and implement \ek to provide cloud-like coordination for MEC systems. It provides naming, network management, application coordination, and security to distributed edge computing applications. It maintains an edge cluster among devices and intelligently stores its data on a group of replicas to guard against node failure and disconnections. We provide a full-system implementation of EdgeKeeper for Android and Linux platforms. We have integrated EdgeKeeper with existing MEC applications and performed real-world performance evaluations in a wide-area search and rescue operation conducted by first responders, which proves it to be lightweight and suitable for mobile devices. 
\end{abstract}

\section{Introduction}


\begin{figure}[t]
\centering
\includegraphics[width=0.8\linewidth]{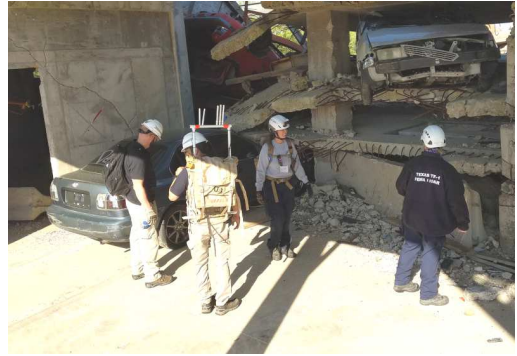}
\caption{A team of first responders, equipped with handheld devices and on body cameras/sensors is deployed for a search and rescue mission after a disaster. The carried mobile devices form a mobile edge that can be used for sharing computation resources. In the absence of network infrastructure, a manpack equipped with LTE and WiFi capabilities provides the required communication backbone for distributed computation at the edge.}
\label{fig:disaster_city_deployment}
\vspace{-5mm}
\end{figure}

Over the last decade, advancements in handheld devices hardware coupled with data rate enhancements in radio access technologies have led to an exponential increase in the number of mobile applications. New applications that generate large amounts of multimedia data, e.g., videos, images, and audio, are continuously being developed. Some of these applications are gaining significant popularity among disaster response and tactical/military teams, e.g., automatic human/scene identification/recognition from video captured through cameras mounted on first responders' helmets or voice assistance~\cite{atak-dhs,atak}. These applications, however, require significant resources for data processing. Traditionally, to process big data by mobile applications, it is necessary to offload processing-intensive jobs to remote cloud servers. In the absence of connectivity to the cloud, as it is the case for disaster response and tactical environments, enabling data processing on mobile devices at the edge becomes a key emerging necessity~\cite{rahimi2014mobile,chen2016efficient,huerta2010virtual,dinh2013survey}. 

Mobile Edge Computing (MEC) enables a paradigm shift in data processing, where jobs are offloaded to the nearest devices instead of sending them to remote cloud servers~\cite{satyanarayanan2017emergence,hu2015mobile,chen2015computation,flores2015mobile}. Instead of treating mobile phones as thin clients, higher computing power allows us to view them as thick clients or effectively thin servers. The MEC scenario becomes more prominent when a group of mobile nodes loses their connection to the Internet, thus to the cloud. This is common in the mission-critical military or first responders networks. A group of mobile devices forms an ad-hoc network but are deprived of high-speed internet connection. In this scenario, as illustrated in Figure~\ref{fig:disaster_city_deployment}, the mobile edge cloud, comprising the crowdsourced smartphones and tablets, becomes the only possibility for big data processing. Currently, a situational awareness mobile app, namely Android Team Awareness Kit (ATAK), is widely used in mission-critical field deployment~\cite{atak,atak-dhs,usbeck2015improving}. It provides the backbone for situational awareness, map data overlay, and tactical analysis to coordinate the troops deployed at the remote field. Although this kind of application provides situational awareness, it does not provide the necessary edge computing framework.

Motivated by a lack of distributed edge computing ecosystem that can run on handheld devices, we developed several applications that suit such dynamic environments. Similar to Apache Hadoop~\cite{hadoop} ecosystem, we developed the \projectName{} ecosystem, which is particularly targeting edge networks formed by handheld devices and deployable manpacks carried by first responders. Many edge computing applications, such as mobile stream processing (MStorm), require a naming and coordination service to distribute orderly execute tasks. Other edge applications such as mobile distributed file system (MDFS) requires resilient storage for critical metadata. In the distributed cloud computing framework, Apache ZooKeeper is widely used for service coordination. However, it requires statically assigned coordination servers and fails to work as soon as the majority of the coordinators get disconnected. Thus, the ZooKeeper at its current state is not suitable for the distributed edge computing at the edge. To fill this gap, we develop a resilient, distributed coordination service for edge computing, called \ek. It is implemented as an application that runs on all the devices in the background and provides resilient coordination for other applications, e.g., device naming, application coordination, edge status monitoring, and authentication for the edge applications. In particular, the following are the main contributions of this paper:

\begin{itemize}[itemsep=0mm]
    \item We design EdgeKeeper, a distributed application coordination service for mobile edge networks.
    \item EdgeKeeper provides a comprehensive API for cleint applications that includes device naming, application coordination, metadata storage, and authentication and edge status monitoring.
    \item EdgeKeeper is designed to ensure resilience of coordination in edge networks and hides all the complexity of edge coordination from applications. 
    \item We provide an implementation of EdgeKeeper for both Linux and Android environments. \footnote{An open-source implementation is available on Github}. 
    \item We conducted real-life deployments with first responders, which prove it to be lightweight and suitable for mobile hand-held devices.
\end{itemize}

The rest of the paper is structured as follows. Section~\ref{sec:background} provides the background and motivation of the current work. The design and implementation of EdgeKeeper are presented in Sections~\ref{sec:design_overview} and~\ref{sec:implementation}, respectively. In Section~\ref{sec:evaluation} we evaluate the performance of EdgeKeeper. Finally, Section~\ref{sec:conclusion} concludes our paper. 

\section{Background and Motivation}
\label{sec:background}

\begin{figure}
	\centering
	\includegraphics[width=\linewidth]{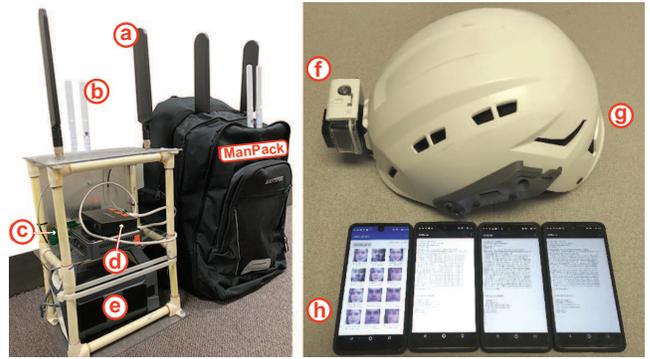}
	\caption{\projectName{} hardware components: a) LTE antenna, b) WiFi AP, c) LTE eNB, d) Intel NUC that runs LTE EPC and HPC, e) Battery, f) Body camera, g) Helmet of first responder, h) Handheld Android phones.}
	\label{fig:hardware}
\end{figure}

\begin{table}
\small
\caption{Abbreviations used in the paper}
\label{tab:abbreviation}
\begin{tabular}{ll}\hline
Abbreviation & Definition \\\hline
ATAK                      & Android Team Awareness Kit           \\
CA                        & Certification Authority              \\
DHCP                      & Dynamic Host Cotrol Protocol         \\
DNS                       & Domain Namming System                \\
DTN                       & Delay Tolerant Network               \\
eNB                       & Evolved Node B                       \\
EPC                       & Evolved Packet Core                  \\
ETx                       & Expected Transmission Count          \\
GNS                       & Global Name Service                  \\
GUID                      & Globally Unique Identifier           \\
HDFS                      & Hadoop Distributed File System       \\
HPC                       & High Performance Computing device    \\
MDFS                      & Mobile Distributed File System       \\
MEC                       & Mobile Edge Computing                \\
MMR                       & Mobile MapReduce                     \\
MStorm                    & Mobile Stream processing application \\
PDR                       & Packet Dropping Rate                 \\
RSock                     & Resilient Socket middleware          \\
RTT                       & Round Trip Delay                    \\\hline
\end{tabular}
\end{table}

Before going into the design of \ek{}, in this section, we discuss the backgrounds of edge network ecosystems and the motivation for \ek{}. First, we present the hardware architecture of \projectName, an edge computing ecosystem designed for disaster response teams. Then, we present the software architecture of \projectName{} and its need for resilient application coordination (similar to what is needed in the cloud). We use many abbreviations throughout the paper, and for better readability, we enlist all the abbreviations in Table \ref{tab:abbreviation}.

\begin{figure*}
	\centering
	\begin{subfigure}[c]{0.30\linewidth}
		\includegraphics[width=\linewidth]{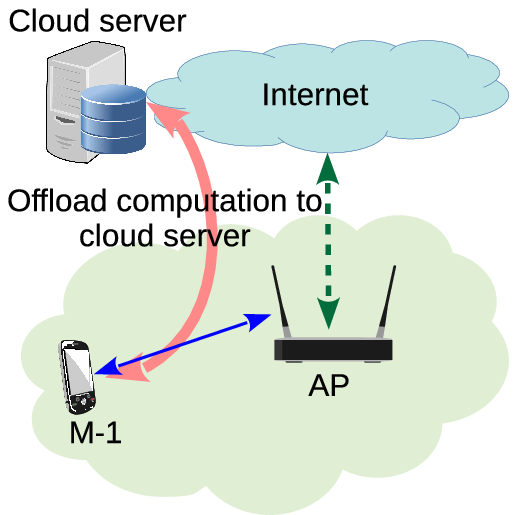}
		\caption{Cloud computing framework}
		\label{fig:arch_cloud}
	\end{subfigure}
	\hspace{0.03\linewidth}
	\begin{subfigure}[c]{0.60\linewidth}
		\includegraphics[width=\linewidth]{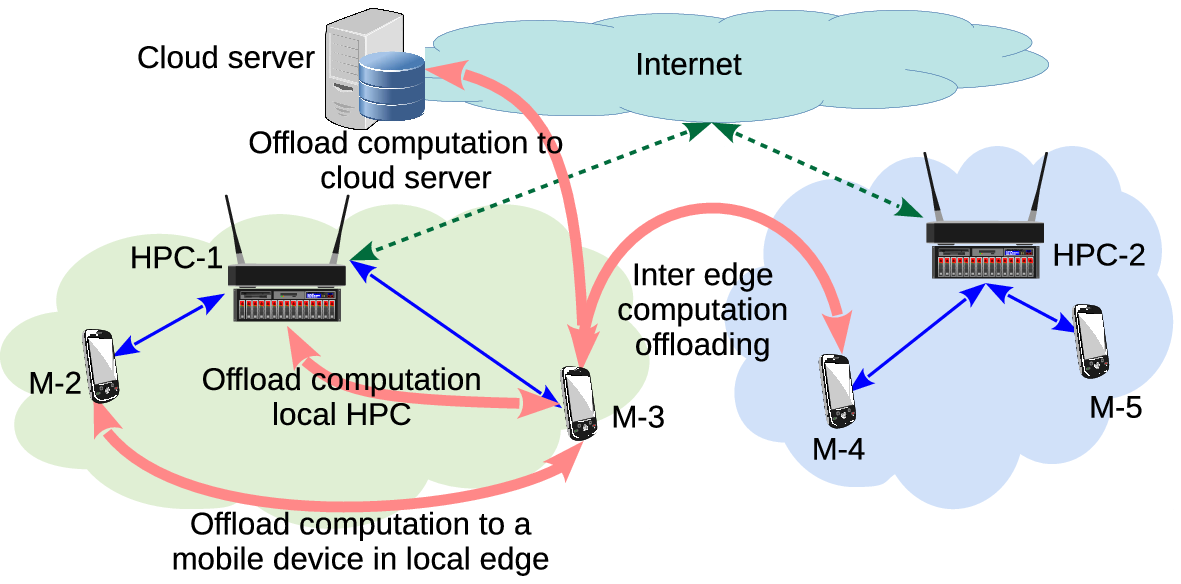}
		\caption{Edge computing framework: }
		\label{fig:arch_edge}
	\end{subfigure}
	\vspace{-10pt}
	\caption{Network architecture for cloud computing vs edge computing framework}
\end{figure*}
\subsection{\projectName{} Hardware Architecture}
A group of first responders carries out its search and rescue mission in a disaster response scenario, assisted by handheld devices, on-body cameras, and other sensors. As the cellular wireless infrastructure is usually unavailable, these teams carry a deployable wireless communication system, which, typically, consists of the following: 1) mobile devices equipped with LTE and WiFi wireless capabilities, 2) LTE eNodeB (i.e., an LTE access point), 3) WiFi access point(s), 4) and High-performance Computing (HPC) device(s). Besides, the edge network can connect to cloud servers through the Internet. \projectName{} hardware is shown in Figure~\ref{fig:hardware}. A lightweight manpack provides LTE and WiFi connectivity to mobile devices. The manpack comprises a Baicells eNodeB and a Ubiquiti WiFi access point. The manpack also contains an Intel NUC as a high-performance computing (HPC) resource. The EPC, MME, and PGW functionalities of the LTE are managed by an open-source NextEPC~\cite{nextepc} running on the NUC. A Ubiquiti EdgeRouter connects all components using Gigabyte ethernet. 



\begin{figure*}
	\centering
	\begin{subfigure}[c]{0.4\linewidth}
		\includegraphics[width=\linewidth]{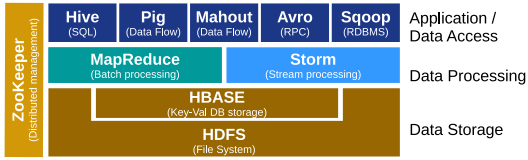}
		\caption{Apache Big Data processing Ecosystem}
		\label{fig:hadoop_ecosystem}
	\end{subfigure}
	\hspace{0.05\linewidth}
	\begin{subfigure}[c]{0.4\linewidth}
		\centering
		\includegraphics[width=\linewidth]{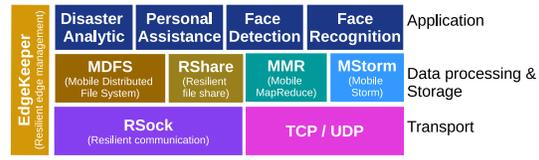}
		\caption{\projectName{} edge computing Ecosystem}
		\label{fig:distressnet_ecosystem}
	\end{subfigure}
	\vspace{-10pt}
	\caption{Software architecture for Apache Hadoop and \projectName{} ecosystems}
	\vspace{-10pt}
\end{figure*}

In the traditional cloud computing paradigm (Figure~\ref{fig:arch_cloud}), a resource-constrained mobile device offloads the computation tasks to a remote server in the cloud. When a mobile node loses connectivity to the cloud, it fails to perform the required computation. In \projectName, as can be seen in Figure~\ref{fig:arch_edge}, multiple mobile nodes form an edge network where mobile devices can offload tasks to nearby mobile devices as well as the cloud server when it is available.

\subsection{\projectName{} Software architecture}

Apache Big Data ecosystem has become synonymous with big data processing in the cloud. To attain a quick response, Hadoop \cite{hadoop} splits a submitted job into several small tasks and executes these tasks parallel on multiple servers, thereby reducing the delay associated with a sequential execution. Furthermore, to enhance the reliability and efficiency of data storage, Hadoop Distributed File System (HDFS) is used, which replicates the data according to its demand and reliability requirements \cite{hdfs}. Figure~\ref{fig:hadoop_ecosystem} depicts the different components associated with the Apache big data processing ecosystem. HDFS has two kinds of servers, namely, Name node and Data node. Name nodes are primary nodes that store metadata of a file. Metadata contains the number of blocks of a file and where the replicas of each block are stored. Data nodes are the servers that actually store the data blocks. MapReduce is a parallel processing framework that consists of two steps. The first step, Map(), takes care of filtering and organizing data in groups, and then in parallel, it processes each group.  In the second phase, Reduce() summarizes the outputs generated by Map() process and combines them to a smaller set of tuples as the final output. The storm is a distributed processing framework that, unlike MapReduce, process data as a stream (i.e., real-time). ZooKeeper~\cite{hunt2010ZooKeeper} is a coordination service that provides coordination to all components of the Apache ecosystem. For better reliability, ZooKeeper replicates data over multiple nodes and maintains consensus among them.  

The Apache ecosystem is not designed to run on mobile devices and performs poorly with servers' mobility. To tackle this issue, we developed similar components to run on mobile devices. Figure~\ref{fig:distressnet_ecosystem}  provides an overview of the \projectName{} ecosystem. We developed parallel processing software mobile MapReduce (MMR) and mobile Storm (MStorm) that run on mobile devices and provide the same functionality as MapReduce and Storm. To tackle the device mobility and frequent disconnections, we developed a resilient socket (RSock), which abstracts data delivery for upper-layer applications in wireless networks with diverse connectivity. We also develop a mobile distributed file system (MDFS) to store large data over mobile devices and attain high reliability using erasure coding. 




\subsection{Motivation for Edge Coordination}
All \projectName{} applications need a reliable and resilient coordination. Unfortunately, conventional coordination services such as ZooKeeper fail to operate in mobile edge environments because of the frequent node and link failure. In ZooKeeper, the server node configuration, which runs the consensus, must be static. If the ZooKeeper ensemble loses the majority of the server nodes, the whole ensemble fails to work. When some nodes leave the edge network, new nodes should be dynamically chosen to participate in the consensus and reconfigure the edge network. The upper layer applications should continue their operation. The following are the main requirements for distributed edge computation. 



\subsubsection{Device Naming}
Most of the applications, such as MMR, use domain naming service (DNS) to abstract the physical IP address for communication with a target device. The conventional hierarchical DNS based naming service fails to handle intermittent network disconnections and high node mobility. As a group of nodes moves away from the DNS server and forms an edge network, the new network can not use the naming service. Moreover, in the mobile edge network, the nodes connect and disconnect very frequently and change their IP addresses. The service should be able to provide name resolution in all available network scenarios.



\subsubsection{Application Discovery and Coordination}
In addition to the naming service, distributed applications also require a coordination service, which provides discovering service providers, data synchronization, group configuration, leader election, status monitoring, critical section handling, queuing, etc.  The service discovery should provide a global view of the available servers for a service, and when possible, provide a subset of nearby servers. Now, a node can move from one edge network to another edge network or may become isolated. This node might require some service running on another device, or this node was offering computing service to other nodes. The coordination service should accommodate the separation from the former edge and association to the new edge without requiring upper-layer applications to handle mobility-related complexities.


\subsubsection{Resilient Metadata Storage}  
Applications such as MDFS store file metadata in resilient storage. Conventionally, these applications store a large amount of data over multiple devices. In contrast, the file metadata (where file fragments are stored) is stored at a master device called name node. The metadata must be stored over multiple devices, ensuring that the metadata could be retrieved even if some of the nodes get disconnected.

\subsubsection{Authentication and Authorization}
A distributed computing framework requires devices or end-users to be authenticated and authorized before offering any service. In cloud computing, several authentication protocols, such as Kerberos, certificates are widely used. These services fail to work when disconnected from the authorization entity. In an edge network, A node might be authorized in an edge network for accessing services. Now, when the node moves to a new edge network that is disconnected from the Internet and the former edge network, the node can not be authenticated or authorized by the ensuing cluster. Thus we need a framework where devices can be authenticated and authorized by an edge network autonomously in connected and disconnected scenarios.

\subsubsection{Monitor Edge Status} 
The client applications require the edge network knowledge (such as link qualities between peer devices, device battery percentage, device load) to make intelligent decisions on their computation or data offloading. Typically these applications assess the edge network individually, resulting in havoc congestion. To reduce the overhead, we need a single service in an edge network to provide a comprehensive view of the edge to client applications. Most of the standard deployable networks consist of multiple network middleboxes such as firewalls~\cite{sherry2012making,lukovszki2016s}. These middleboxes disrupt the conventional ad hoc link maintenance and routing protocols. Thus, the \ek{} needs to provide an edge status monitoring service that can work with the standard deployable networks. 

\begin{figure*}[t]
	\centering
	\includegraphics[width=0.75\linewidth]{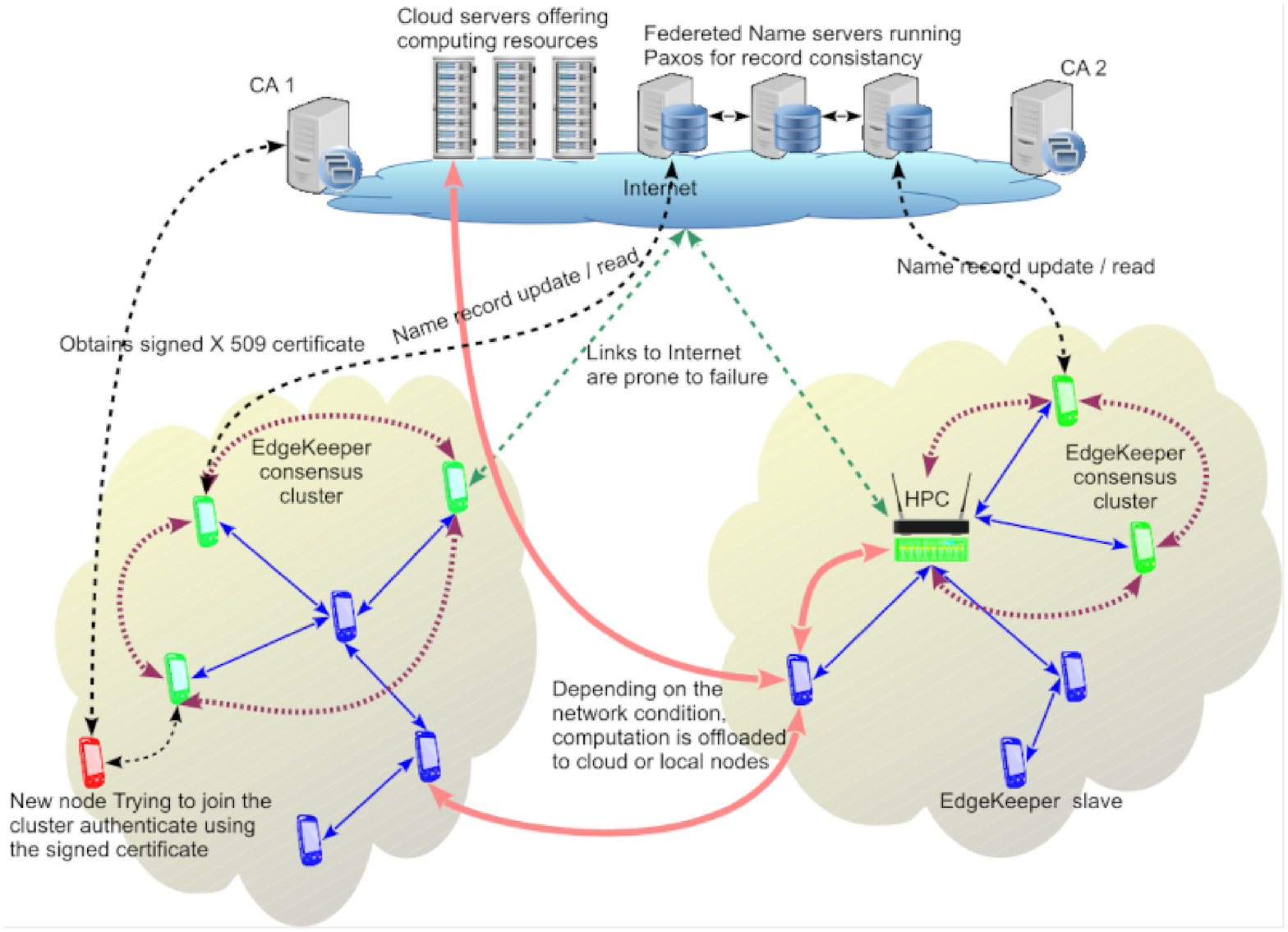}
	\caption{\ek{} software architecture.}
	\label{fig:network_component}
\end{figure*}

\section{EdgeKeeper Design}
\label{sec:design_overview}

As discussed in the earlier section, we need a coordination service for mobile edge networks to provide resilient device naming, service coordination, metadata storage, authentication, authorization, and edge status monitoring. This section offers the design of a resilient and lightweight coordination service, namely \ek{}. Figure~\ref{fig:network_component} shows the main components of \ek. It runs on all devices as background processes. Instead of running one \ek{} at a central device and storing all data at a single node, data is replicated over multiple devices to tackle link failure. The devices that store the data with consistency are called \ek{} replicas and provide \ek{} functionality to the slaves. The role of replica and slave is chosen dynamically, depending on the network status. In the following sections, we describe services offered to the client application.




\subsection{Device Naming} 
\ek{} provides resilient device naming for edge network, which can get disconnected. When connected to the Internet, it must provide a coherent name resolution for a global scale. \ek{} uses Global Naming Service (GNS)~\cite{sharma2014global}, which employs multiple name servers to deal with high name resolution rates across the globe. Each name record is associated with a primary key: a globally unique identifier (GUID). Unfortunately, if a GNS server gets disconnected from the federated group of GNS servers, it fails to provide services. GNS servers fail when disconnected from the global network. \ek{} uses a local cache mechanism to store the name records at the edge. The name record updates are committed at the local cache and lazily updated to the GNS server whenever the edge network gets connected to the Internet. The cache is maintained by the \ek-replicas, who run consensus among themselves for consistency. The \ek{} running at every node triggers updates whenever a device changes its IP address. The updated IP is stored at the local \ek{} cluster, as well as updated to federated GNS if the node can reach GNS. 
\ek{} provides the following API for device naming:
\begin{verbatim}
getOwnGUID()
getOwnAccountName()
getIPbyGUID( guid)
getGUIDbyIP(ip)
getGUIDbyAccountName(accountName)
getAccountNamebyGUID(guid)
\end{verbatim}

The \ek{} also serves the conventional DNS based name to IP translation.  In the typical case, the DNS translator should reside at the HPC as it runs the DHCP for LTE and WiFi. Upon receiving a DNS query, this translator tries to resolve the name by two methods: 1) checking with local GUID record in the cluster for IP translation; and 2) forwarding the query to one of the GNS servers. The DNS server will return the value whichever query returns first. When GNS servers are unreachable, the DNS query for hosts residing in the local edge succeeds. 

\subsection{Service Discovery and Coordination} \label{sec:service-discovery}
\ek{} uses GUID records for service discovery. If a device offers some service and wants it to be discovered by other nodes in the network, the service name and the role (e.g., server, client, etc.) are mentioned in the GUID record. Each GUID record contains an associative array of key-value pairs, as given below:

\begin{verbatim}
{GUID: <own GUID>, 
alias: <host name for DNS>,
netaddress: [<node's IP-1> , <node's Ip-2>], 
<application-name 1>: <application-role 1>,
<application-name 2>: <application-role 2>,
last-update: <node's system time> }
\end{verbatim}

 Any node who wants to find a list of nodes offering a particular service will query to retrieve a list of GUIDs, which contains the key-value pair as \verb|service: role|. Bellow is the client API for application discovery:

\begin{verbatim}
addService(ownService, ownDuty)
removeService(targetService)
getPeerGUIDs(targetService, targetDuty)
getZooKeeperConnectionString()
\end{verbatim}

\subsection{Resilient MetaData Storage} \label{sec:metadata-storage}
\ek{} also provides resilient metadata storage to client applications. To eliminate the problem of a single point of failure, it uses ZooKeeper, an open-source consensus implementation. In our design, the replica nodes run the ZooKeeper server process and participate in the consensus. When the replica cluster cannot reach consensus due to the majority of ZooKeeper server failure, new nodes are chosen dynamically to be replicas. Note that, maintaining consensus over wireless links is costly as it involves many message exchanges between replicas. Thus, an intelligent decision should be made for the number of replicas. Following are API for metadata storage:  

\begin{verbatim}
putMetadata(metadata)
getMetadata(filePathMDFS)
\end{verbatim}

\subsection{Authentication and Authorization}
For authentication purpose, \ek{} uses X509 certificate-based authentication~\cite{housley2002internet,cooper2008internet}. Each device maintains a public-private key pair for authentication purposes. A GUID is a {\bf self-certifying identifier} as it is derived by a one-way hash function (known universally) from the user's public-key. A bilateral challenge-response could be used to authenticate if a node claiming to be actually the GUID owner. Suppose a node A wants to authenticate another node B. A sends a random nonce $n$ to B. B replies with B's public key and encrypted cipher of the nonce using B's private key. A verifies whether B's GUID is derived from the public key sent by B, and the received encrypted cipher can be decrypted using the public key of B. If the decrypted cipher matches the sent nonce, A can certainly verify that the node actually poses the private-key of B and, hence, is the rightful owner of B's GUID.

We understood that by using GUID, a node could verify if another node actually possesses the private key corresponding to the claimed GUID. However, an adversary can create a certificate using OpenSSL and derive the GUID corresponding to it. The bilateral challenge-response is not sufficient to verify whether this entity is authorized to access resources in the network. In our proposed architecture, a certifying authority (CA) at each organization creates client certificates and sign them using the CA's private key. The CA’s public certificate (.pem) is stored at the TrustStore of the federated GNS servers.  When a new node tries to create a GNS account, it provides the signed client certificate. Since the GNS server already contains the public certificate of the CA, it can verify the client. Multiple CAs can be imported to the trust store of GNS servers; thus, multiple entities can have their own CA and provide them to GNS. 
 
 The client credential (public-private key pair and a signed certificate from CA) is stored in a p12 file. A p12 file contains:  
 1) A public certificate which contains user credentials (Identity/name, organization, etc.), the public key and the digital signature from CA;
 2) Public certificate of the CA; and
 3) The private key corresponding to the public key of the user.
 The p12 file is password protected such that the private key is not disclosed to an unauthorized entity.
 
 The local cluster initially does not possess any CA certificate. In our architecture, to maintain a similar authentication mechanism as GNS, \ek{} only accept an update or query request from a device if the requesting entity is verified through the public key. Every time a node connects to a \ek{} cluster, it checks whether the node is authenticated or not. If not authenticated, then the authentication process is initiated. The new node sends the authentication request with its own GUID and the signed certificate from the CA. If CA’s public certificate is already stored in the trust store, the node is validated, and node's certificate is stored in the TrustStore. If CA’s certificate is not found in the TrustStore, the \ek{} requests the GNS server to retrieve the public key. If the public key from the GNS and the new node is identical, the node is authenticated, and the node's certificate is stored in the TrustStore. In addition, the certificate of the CA is also added to the TrustStore. We assume that if a node is authenticated by GNS, that means the entity is trustworthy. 
 
When a new node sends an authorization request, if the \ek{} can not reach the federated GNS servers and the public key of the new slave's CA is not stored in the TrustStore, the new slave can not be authorized to the network. In this situation, we may allow any of the already authenticated slave to request new slave authorization. The new slave will present its signed certificate to already authenticated user through Bluetooth or QR code, or other means. An authenticated slave would request the \ek-replica to put the certificate of a new slave in the TrustStore and authorize the user. Also, we can restrict who can request new slave authorization. For example, we can restrict that only the slaves from the same organization of the master (the same CA signs their certificates) can request for authorizing a new user during disconnection.

\subsection{Monitor Edge Status}
At the edge, the applications need two kinds of statuses: 1) the network topology, and 2)  devices status. \ek{} provides the two services as follows. 

\subsubsection{Discovering Network topology}
\ek{} runs a topology discovery service where each device periodically pings other devices in the network to determine the device-to-device link quality. Suppose there are multiple links available between a pair of devices (LTE and WiFi links between A and B). In that case, the topology discovery service maintains separate link qualities for those links. 
\ek{} maintains a graph using the JGraphT library \cite{jgrapht}. Each node maintains the link qualities to its immediate neighbors; when there are multiple parallel links possible, it maintains the parallel links. If any other node can be reached by the neighbor, \ek{} adds a link between the neighbor node and the remote node. Periodically each node calculates the optimal distance from itself to all the destination and broadcasts this distance vector to its neighbors. Thereby, the whole network is seen as a two-hop network from all nodes. 
Each device periodically sends a UDP packet containing its distance vector to individual neighbors using unicast IP addresses. Each node maintains the GUIDs of its neighbors and the IP addresses to send the periodic messages. There are several metrics available to measure link quality, such as bandwidth, roundtrip delay (RTT), packet dropping rate (PDR), Expected transmission count (ETx), etc. Measuring the bandwidth over the wireless link is difficult because it requires periodically probing a link, and all other traffic has to be stopped. Thus, we refrain from measuring the bandwidth. For link quality, we measure RTT and ETx. RTT is measured as the time taken to get a reply for a periodic ping message. For each link, \ek{} stores an average value of the measured RTTs using an exponential moving average method. Let's consider the calculation of ETx between a link between A and B. ETx of a link between A and B can be calculated as: $$ETx_{AB}=\frac{1}{(1-PDR_{A\to B})(1-PDR_{B \to A})}$$, where $PDR_{A\to B}$ is the packet dropping the rate from A to B and  $PDR_{B \to A}$ is the packet dropping the rate from B to A. The simple notion behind this formula is that ETx is the expected number of transmission required for a packet to be successfully transmitted from A to b  and its corresponding acknowledgment to be received at A.

We have also observed that deployable systems sit behind network address translators. Any node inside a deployable edge network can reach a node residing in the cloud as the cloud nodes expose their IPs to the global view. However, cloud nodes can not initiate data delivery to nodes sitting behind the NAT. Yet, if a TCP connection is initiated from an edge network to the cloud server, the cloud server can reply to the edge network through the established session. When applications such as MStorm want to offload computation to the cloud server, it needs to know the link quality to the cloud server as well. To bypass this issue, we allow the manpack (HPC) node to assess the link quality of the link between itself and the cloud server (for our case, the \ek{} running in the cloud) using periodic pings through TCP sessions. The manpack then includes this link quality information when sharing the distance vector table with its neighbors. All other nodes on the edge network add this link in their topology graph. The following are the APIs for obtaining topology information. 

\begin{verbatim}
getNetworkInfo()
getAllLocalGUID ()
\end{verbatim}

\subsubsection{Device and Application Health monitoring}
Edge applications require the status of peer devices in terms of the number of functioning processors, available memory, remaining battery, and available storage, etc. The \ek{} running on each device periodically measures the device status and reports it to the local \ek-master. Any edge application can obtain a device's status through the resident \ek{}, which pulls it from the \ek-master. Applications can also report application-specific status, such as queue length, processing latency. The following are the APIs for device monitoring:
\begin{verbatim}
putAppStatus(appName, appStatus)
getAppStatus(targetGUID, appName)
getDeviceStatus(targetGUID)
\end{verbatim}

\section{EdgeKeeper Implementation}
\label{sec:implementation}
In the previous section, we discussed the design of \ek{}. This section describes the implementation of with specific detail on how an edge network is formed and how nodes' joining and departing effect the edge network.  \ek{} runs on all devices: as a daemon in Unix and as a background service in Android. All client applications such as MStorm, RSock, Madoop, and MDFS interact with the \ek{} service running in the local device. The client application uses a Java client library, which communicates with \ek{} process through JSON based RPC over a local TCP socket. Nest, we discuss how an edge network is formed. 
\begin{figure}
	\centering
	\includegraphics[width=\linewidth]{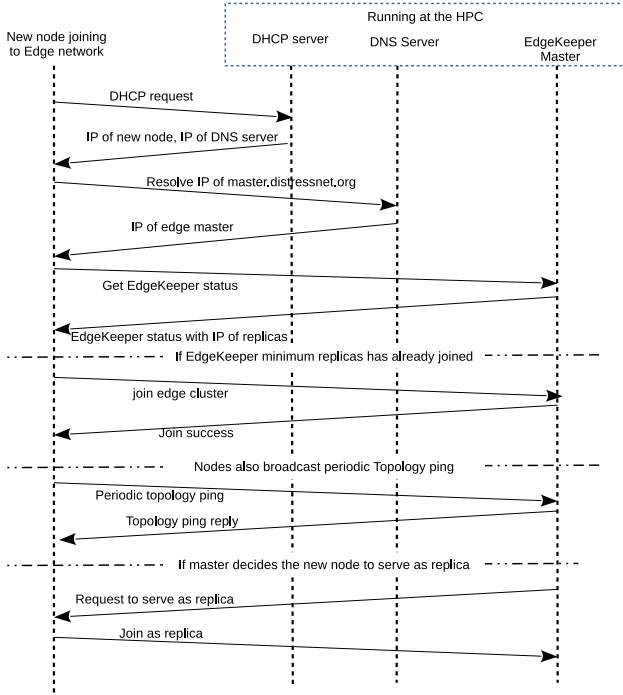}
	\caption{Time diagram of message exchanges when a new node joins an edge network}
	\label{fig:timing_diagram}
	\vspace{-2mm}
\end{figure}

\subsection{Discovering Nodes using \ek{}  Master} \label{sec:master-discovery}




As \ek{} cluster is formed using distributed processing entities running over multiple nodes, the nodes need to discover other nodes in the cluster. Discovering neighbors in a LAN is a well-studied topic. There exist many neighbor discovery protocols such as  ZeroConf  Link-local multicast name resolution (LLMNR) \cite{aboba2007link}, Multicast DNS (mDNS) \cite{sundaresan2016mdns}, DNS Service Discovery (DNS-SD) \cite{cheshire2013dns}, etc.  
There are several widely used implementation of zero-configuration service discovery protocols such as Bonjour and Avahi \cite{Avahi,bonjore}, which works on OsX and Linux, respectively. These protocols provide a general method to discover services on a LAN and only works within a single broadcast domain. An example includes finding a printer, file share, iPhotos, etc. After analyzing several deployable networks, we found that enabling the aforementioned neighbor discovery protocols on a conventional deployable network requires a lot of reconfiguration. To keep the network reconfiguration minimal, we developed a lightweight method through which a new node can find other nodes in the network. 

We designate a node as a gateway where all new nodes should send their joining message first. We call this gateway node as \ek-master. The node joining procedure is depicted in Figure~\ref{fig:timing_diagram}. The first step in this process is finding the master node. After joining a wireless network using WiFi or LTE, the new node obtains the IP address from a DHCP server. In the network, the \ek-master runs the DNS server, and the DHCP reply contains the master's IP in the DNS field. After joining a new network, the new node uses a special hostname  \texttt{master.anonymous.org} to find the IP address of the master. 

The second step in this process is to start the topology discovery. The new node sends a topology ping message to the master. The ping message contains the new node's GUID, IPs, and a sequence number. Upon receiving the ping message, the master replies to the ping with its GUID and IPs. The master also ads this device to its topology graph. Periodically, the master sends the distance vector table to all neighbors. Upon receiving the periodic distance vector table from the master, the new node obtains the information about other nodes in the network. Afterward, the node starts pining the network and updates the topology graph accordingly. 

Now, for the new node's \ek{} to access the GUID records maintained by \ek{} replicas, the new node needs to access the \ek-replicas. The new node sends a request to the master for the \ek-status. The master replies about the current status of the replicas and the IP address of the replicas. If the replica status confirms that the minimum number of replicas are present, the ZooKeeper client in the new node uses the IP address of the replicas to connect. If the minimum number of replicas for a quorum is not met, the new node waits and fetches the replica status repeatedly. If the master decides to use this node as one of the replicas, the master sends a request to the node to join as one of the replicas. 

\subsection{Selection of \ek-Replicas} \label{sec:replica-selection}

The required number of replicas is stored in a configuration file of the \ek-master. If the replication number is 1, the master chooses itself as the only replica. For multiple replicas, the master periodically checks the network topology and monitors the presence of other nodes. From the topology, it finds the one-hop neighbor suitable to run the replica service. Since the replicas run consensus, the replicas must be reachable with each other without any middle-boxes. For the replication factor of $r$, the master always tries to select $r$ nodes for replicas. If there are less than $r$ suitable nodes, then it tries to assign the maximum possible suitable nodes to serve as replicas. If less than $r/2$ nodes are selected as replicas, the edge status is updated as in \textit{looking} state. If more than $r/2$ nodes are serving as replicas, the state is updated to \textit{formed}. 

\subsection{Resiliency in Replica Failure}\label{sec:resiliency}
We described how a node finds an already established edge network where a master node is already running in the earlier section. Now we need to decide what happens when a node moves out of an edge network. If the departing node is not serving as one of the replicas, there is no change in the edge status. After a predefined time, the node's GUID entry is eliminated from the local \ek-cluster. If the departing node is serving as one of the replicas, a new node should be selected to serve as a replica. The \ek-master periodically checks the topology graph if one of the replicas is disconnected for a certain amount of time. In that case, the node is removed from the set of replicas, and a new node is assigned to serve as a replica. The master informs all the replicas to update the change.  When the replicas are updating their service for this change, the slave might get disconnected for a very short amount of time. Note that, if the majority of the replicas leave an edge network together, then the quorum is broken. The master selects a net set of nodes to serve as the replicas. In this case, previously stored data in the replicas are lost. 

\begin{figure*}[t]
    \centering
	\begin{subfigure}[c]{0.25\linewidth}
		\includegraphics[width=\linewidth]{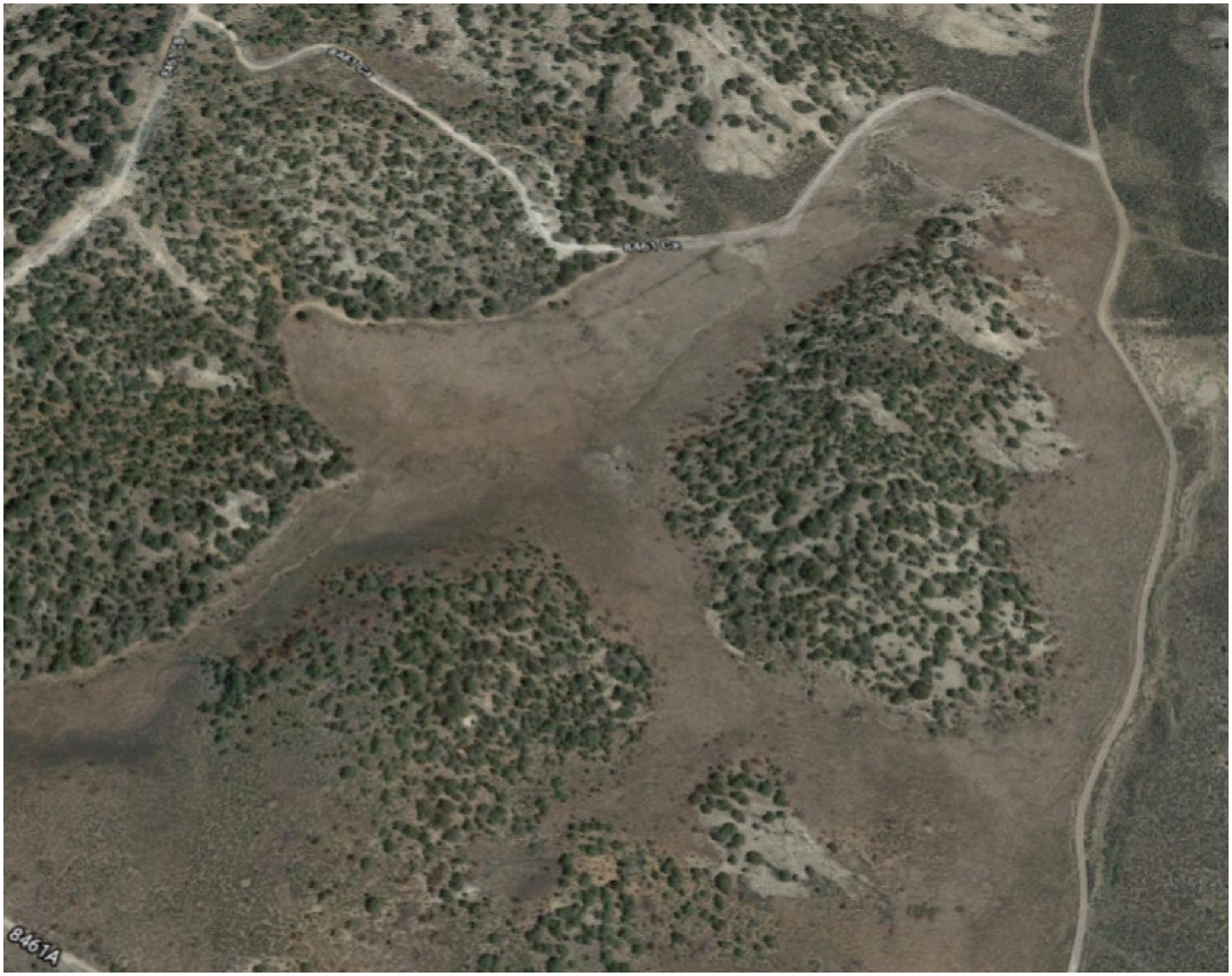}
		\caption{Gypsum, CO open field}
		\label{fig:gypsum_areal}
	\end{subfigure}
	\begin{subfigure}[c]{0.23\linewidth}
		\includegraphics[width=\linewidth]{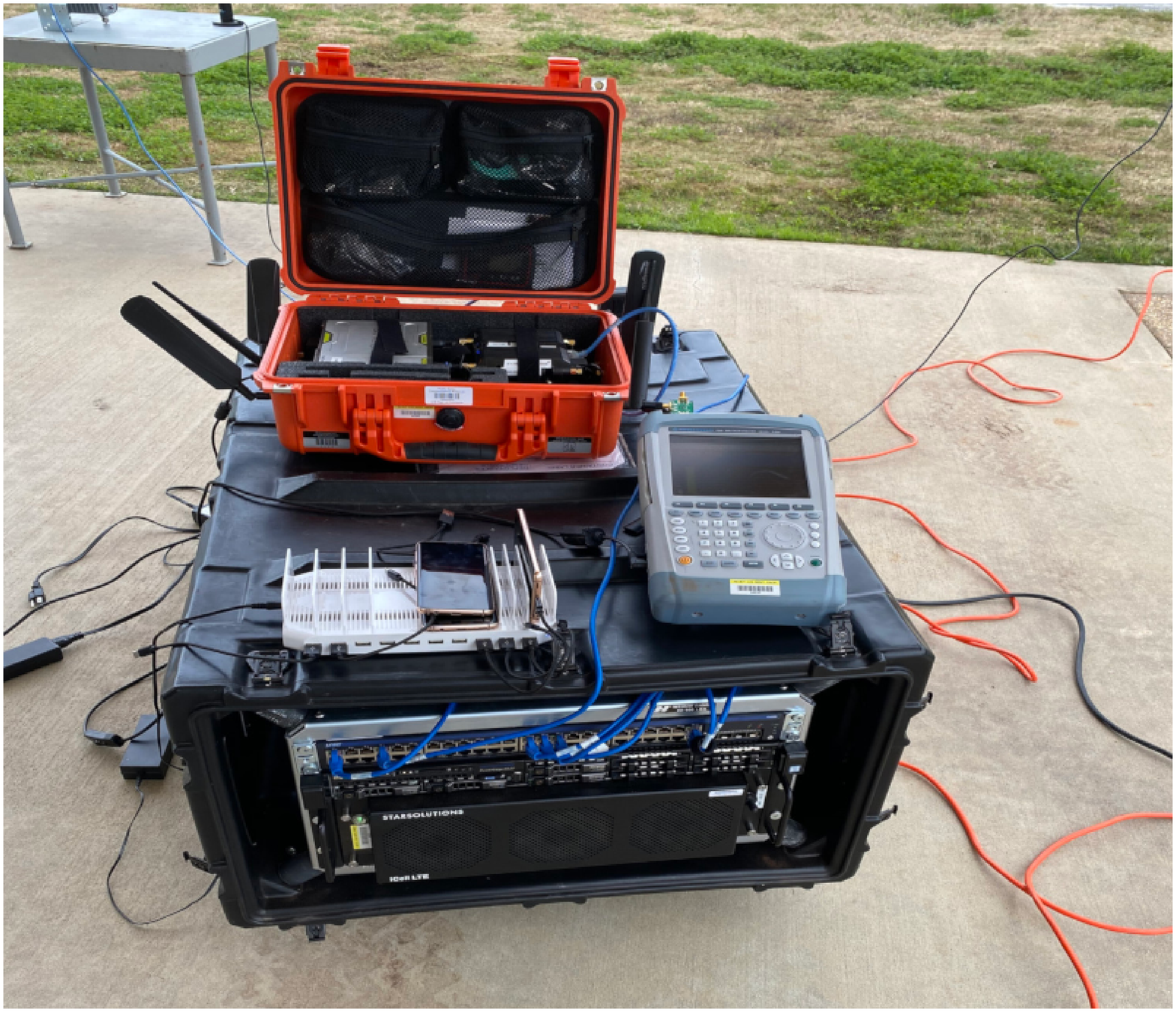}
		\caption{NIST deployable network}
		\label{fig:NIST_deployable}
	\end{subfigure}
	\begin{subfigure}[c]{0.25\linewidth}
		\includegraphics[width=\linewidth]{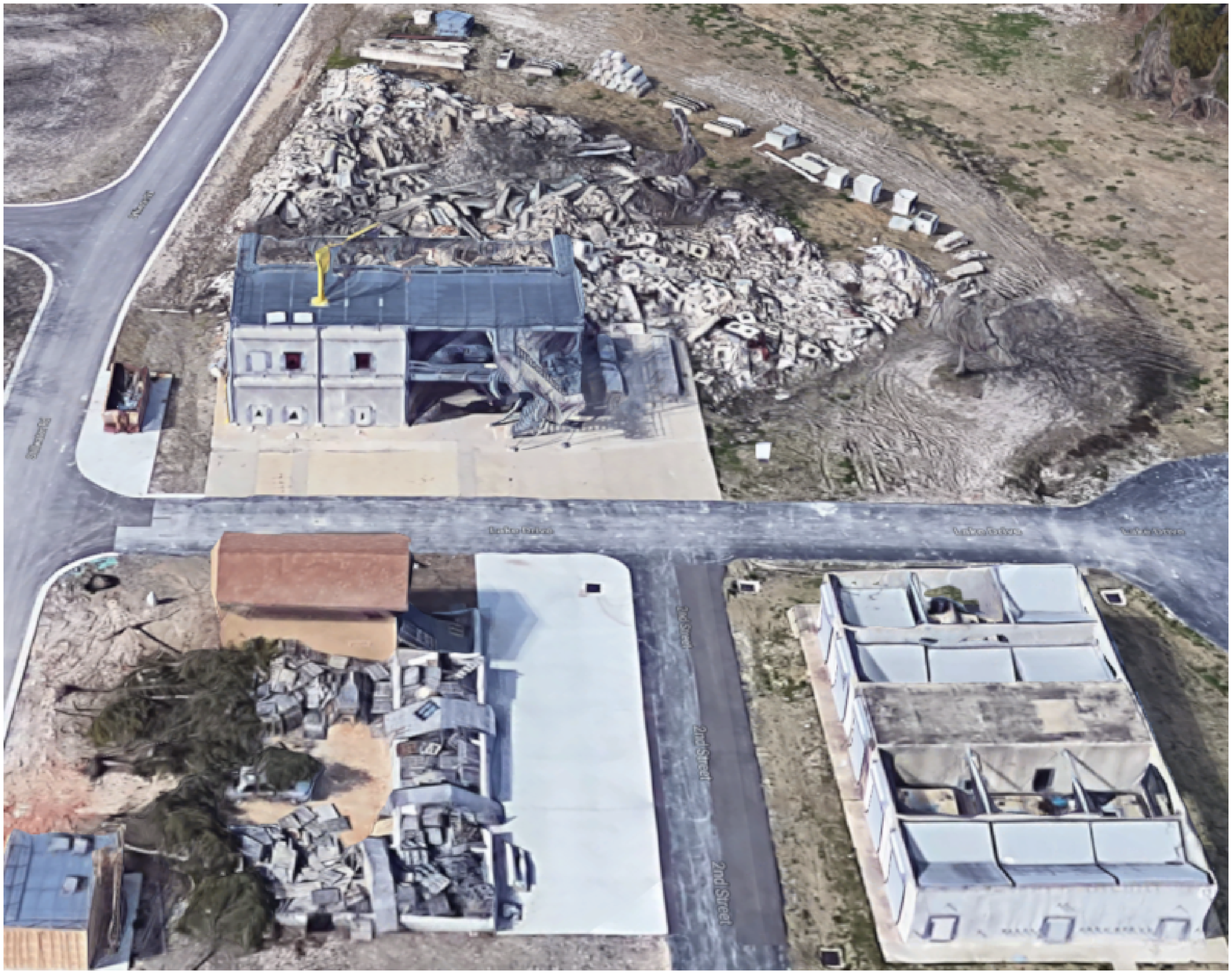}
		\caption{Disaster City, TX}
		\label{fig:disaster_city_aerial}
	\end{subfigure}
	\begin{subfigure}[c]{0.193\linewidth}
		\includegraphics[width=\linewidth]{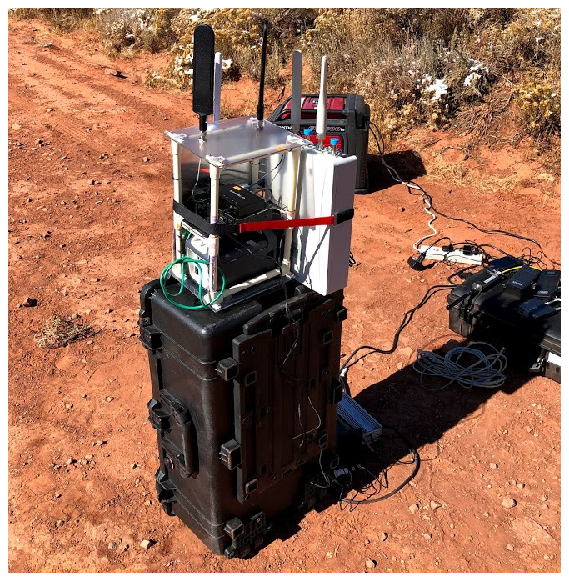}
		\caption{Lightweight Manpack}
		\label{fig:Tamu_deployable}
	\end{subfigure}
	\vspace{-10pt}
	\caption{Real-life deployment.}
	\label{fig:experiment-areal}
 	\vspace{-2mm}
\end{figure*}

\subsection{Forming a New Edge with WiFi-Direct}
If a group of nodes gets disconnected from the HPC node, then those nodes can form a new ad-hoc edge network. A user needs to star WiFi-Direct on one of the nodes. The WiFi-Direct-based network is formed by a group owner, which is usually a mobile phone. Unfortunately, we can not run a DNS server on a mobile phone as it requires 'sudo' privilege. In this case, the new node also tries to treat the default gateway in the network (which is the group owner) as the possible \ek-master. Also, the \ek{} in the WiFi Direct Group Owner needs to be started in the master mode. Once the Group Owner device starts \ek-master, other nodes can join the edge network as described in Section~\ref{sec:master-discovery}.

\subsection{Edge Partitioning and Merging}
Now let's discuss the scenario if two edge networks establish a mesh link to form a single network. In this case, both edges maintain their \ek-replicas separately. The masters share the GUID data and MDFS directory information with each other. Upon receiving this information from a neighbor master, the local master pushes these data to the local replicas. Thus, applications running on edge-1 can discover other nodes running similar services across to edge-2 through service discovery.  


\subsection{Cloud Integration}
\label{sec:cloud-integration}
Many of the client applications, such as MStorm, use high-performance servers available in the cloud. Thus, \ek{} needs to provide a unified interface such that MStorm clients can use the cloud servers when connected to the Internet. To tackle this purpose, \ek{} is run on the cloud servers as a separate cluster. The \ek-master monitors the link quality to the cloud servers. As firewalls blocks UDP traffic from cloud to local edge, \ek-masters initiate a TCP connection to the cloud \ek{} for link quality monitoring through periodic pinging. Client applications use this link quality to determine whether to offload computation to the cloud or use local devices for computation.  The services running in the cloud are discovered by service discovery as the service discovery uses GNS based service discovery.


\section{Evaluation}
\label{sec:evaluation}
In this section, we evaluate \ek{} both in a controlled laboratory environment as well as real-life outdoor deployment. First, we evaluate the internals of \ek{} implementation (described in Section~\ref{sec:implementation}) in a laboratory environment. We use startup time, resiliency in replica configuration, and application overhead as performance metrics. After evaluating the internals of \ek{}, we evaluate quality of client application interface performance of \ek{} (described in Section~\ref{sec:design_overview}). We measure the latency when a  client application makes API calls for service discovery and metadata storage. 

For the indoor experiments, we use the following hardware configuration as depicted in Figure~\ref{fig:Tamu_deployable}. The HPC node (manpack) consists of an Intel NUC (32GB Intel OptaneTM Module, 2TB HDD, 8GB DDR4-2400 SDRAM), Ubiquity EdgeRouterX, BaiCells Nova 227 eNB (2496-2690MHz, maximum power 27dBm), Unify 802.11AC Mesh, powered by a Veracity PointSource Plus PoE injector. For mobile devices we used Essential Ph1 phones (Qualcomm Snapdragon 835 2.45GHz octa-core Kryo 280 CPU Adreno 540 GPU, 4GB RAM, Android 7-9). 


For real-life deployment, we conducted multiple sets of experiments in three different environments. The first set of experiments is carried out in an open mountainous region near Gypsum, CO, (Figure~\ref{fig:gypsum_areal}) where the mobile phones are mostly connected to the LTE network (Figure~\ref{fig:NIST_deployable}) line-of-sight. From this experiment, we are able to understand how far the devices can operate and share computing resources. 
In the second set of experiments, we deployed \ek{} on handheld devices carried out by first responders during a wide-area search and rescue exercise. Each year Disaster City (Figure~\ref{fig:disaster_city_aerial}) organizes an event, Winter Institute, where several first responder divisions (such as FEMA, Police, Fire Fighters, GameWardens, etc.) come together and practice different rescue missions to enhance the collaboration after a disaster. This was a real test opportunity for us to equip the first responders with our products and get feedback as-well-as measure the performances. The response team consisted of 4 responders; one of them carried a specially built lightweight manpack (Figure~\ref{fig:Tamu_deployable}) on the back. One of the responders carried a helmet-mounted camera that recorded video for stream processing. All the responders carried Android phones. The phones connected to the manpack using both LTE and WiFi. We allowed the responders to move freely as per their assigned tasks. Each device runs \ek{} and other client applications such as MStorm, RSock, and MDFS. The third set of experiments is carried out in Christman Airfield, Fort Collins, CO, where the mobile phones are only connected to the LTE network and the F-Lite LTE  eNB(FeatherLite UAV deployable eNB) was airborne on a fixed length short circular flight carried out by a quad-rotor drone. From this experiment, we are able to assess the feasibility of UAV beased deployment of the system and understand how it performs compared to other deployments. The overall evaluation in this setup was limited by the short length of the flight due to the trade-off between weight capacity and battery life of the drone.

\begin{figure}
	\begin{subfigure}[c]{0.48\linewidth}
		\includegraphics[width=\linewidth]{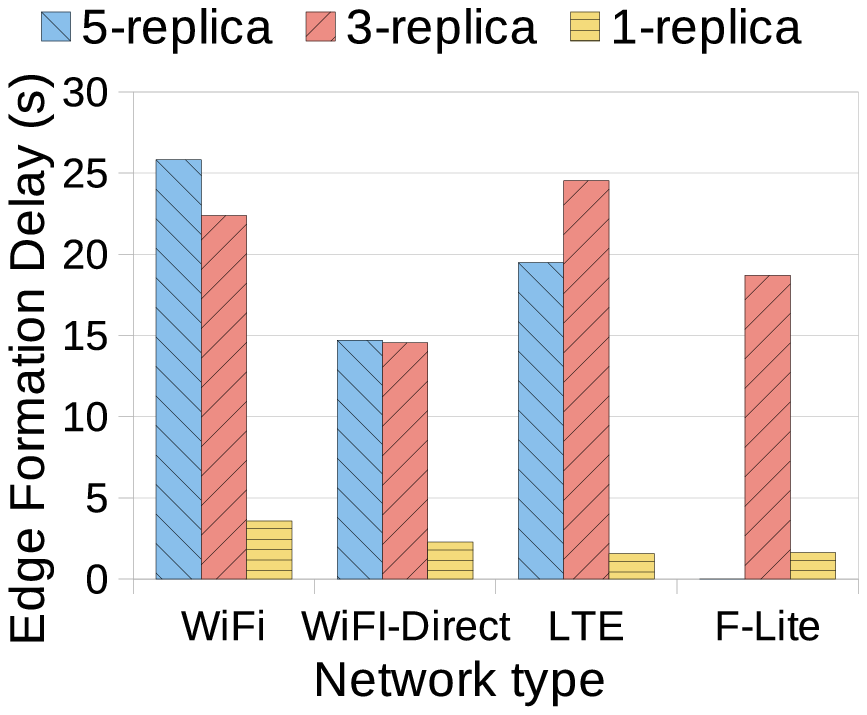}
		\caption{\ek{} startup with at least half of required replicas are present.}
		\label{fig:edge-formation-delay}
	\end{subfigure}
	\begin{subfigure}[c]{0.48\linewidth}
		\includegraphics[width=\linewidth]{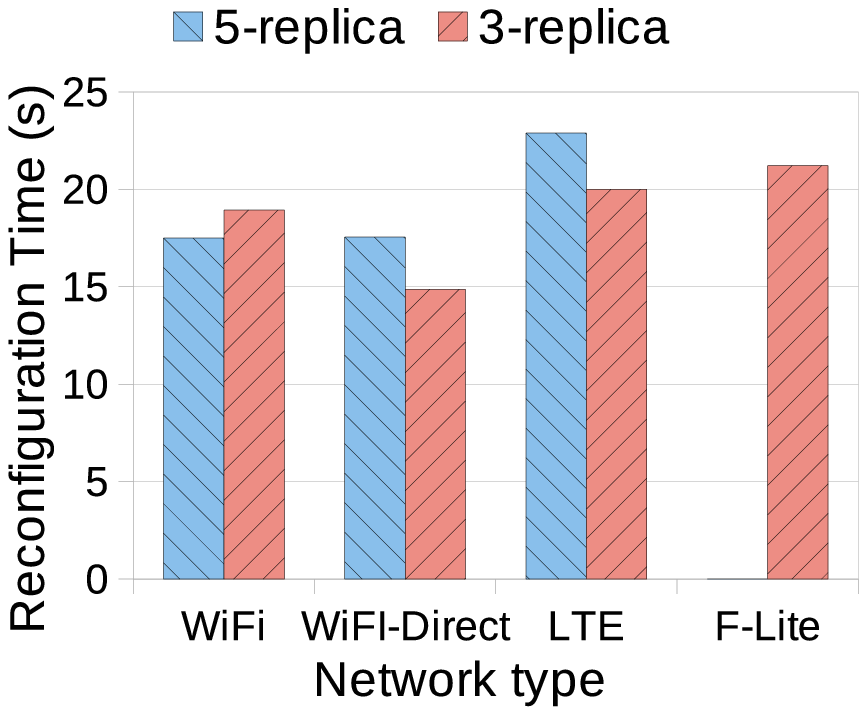}
		\caption{\ek{} replica reconfiguration time after one node joins a partially formed cluster.}
		\label{fig:edge-replica-addition}
	\end{subfigure}
	\vspace{-10pt}
    \caption{\ek{} cluster startup time. This startup time includes the neighbor discovery. The topology ping interval is set to 10s. (5-replica was not conducted for F-Lite setup) }
    \label{fig:edge-formation}
\end{figure}

In the following sections, we present how long an EdgeKeeper takes to start a cluster, resiliency in replicas, overhead on mobile phones, latency for service discovery, and metadata storage performance. 

\subsection{Edge Network Formation and Replica Selection}
As the first step to evaluate \ek{} internals, this section analyzes the edge formation process as described in Section~\ref{sec:replica-selection}. First, we analyze the time taken for starting an \ek{} cluster. We conducted experiments on three different networks: WiFi, WiFi-Direct, and LTE. Figure~\ref{fig:edge-formation-delay} plots the edge formation delay. In all scenarios, 1-replica configuration, \ek{} on all devices, start serving clients quickly after the master is started. However, when we use multiple replica configuration, the \ek{} cluster needs significant bootup time. In these cases, the master needs to discover other nodes in the network and select suitable replicas.  With a replication factor of $r$, the master needs to find more than $r/2$ number of replicas in its topology graph to select replicas. As the periodic ping messages are sent at an interval of 10s, this causes higher delays as can be seen in the plot. In the WiFi and LTE network scenarios, nodes discover the master through DNS resolve, that case higher delay than the WiFi-Direct networks. In WiFi Direct network, the nodes treat the group owner as of the master and directly pings the master without going through the DNS. In most cases, we can see that a cluster is started within 25 s when the topology ping interval is set to 10s. 

Next, we test how long it takes to add one new node as a replica in an already running \ek{} cluster where there is a shortage of replica. In the case of 3 replica scenario, the \ek{} serves clients as long as there are 2-replicas. When another node is added, the master includes it in the replica pool to get closer to the configured replica number. Figure \ref{fig:edge-replica-addition} plots the reconfiguration delay after a node is introduced in the network. In most cases, reconfiguration is also achieved within 25 s.

\begin{figure}
	\begin{subfigure}[c]{0.32\linewidth}
		\includegraphics[width=\linewidth]{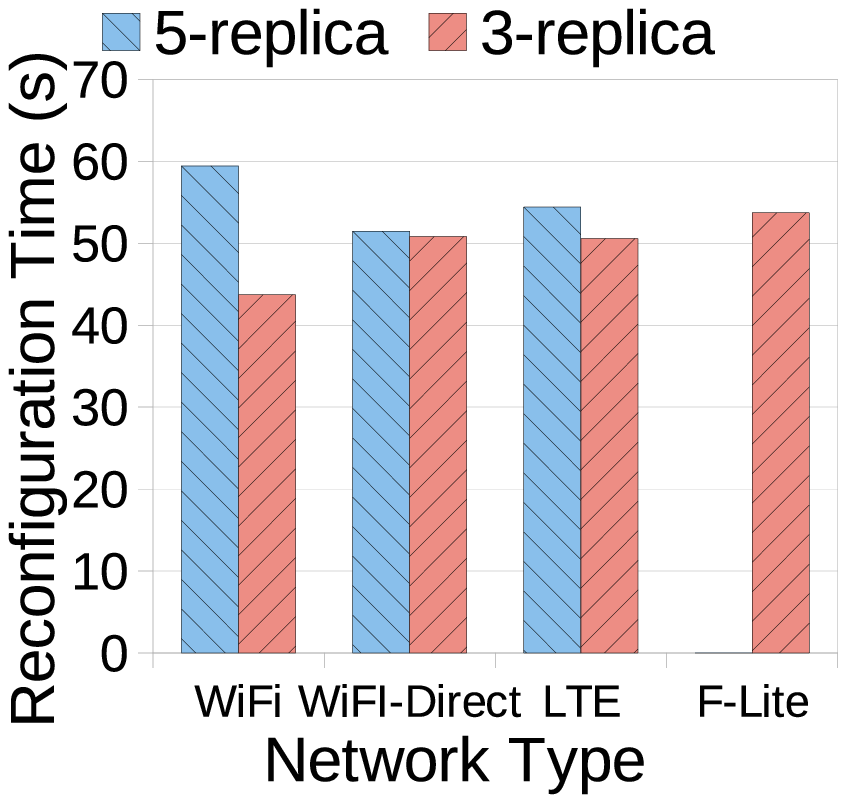}
		\caption{1-replica depart}
		\label{fig:edge-1replica-depurture}
	\end{subfigure}
	\begin{subfigure}[c]{0.32\linewidth}
		\includegraphics[width=\linewidth]{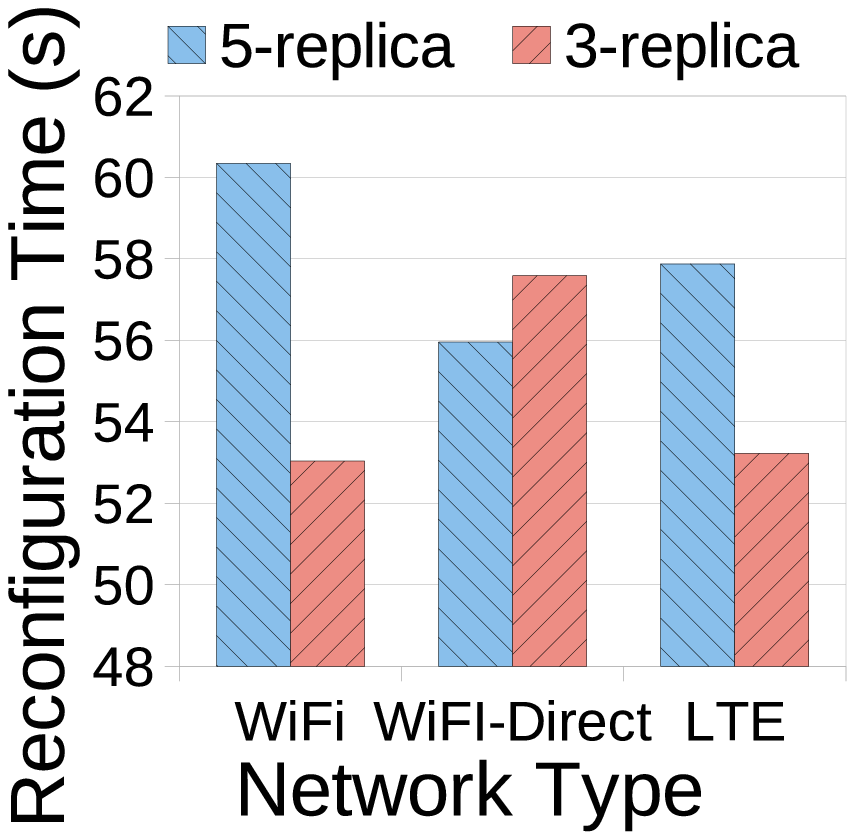}
		\caption{2-replicas depart}
		\label{fig:edge-2replica-depurture}
	\end{subfigure}
	\begin{subfigure}[c]{0.32\linewidth}
		\includegraphics[width=\linewidth]{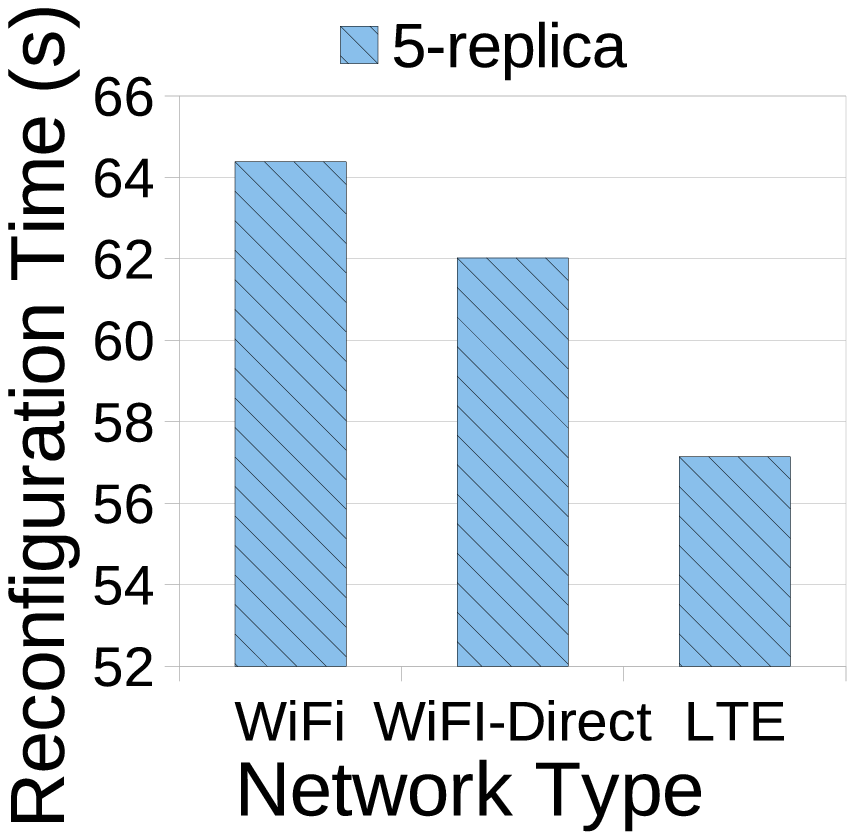}
		\caption{3-replicas depart}
		\label{fig:edge-3replica-depurture}
	\end{subfigure}
	\caption{Edge replica reconfiguration time when some nodes acting as replicas departs the network. (Only 1-replica departure was conducted in F-Lite setup for the limited duration of the flight)}
	\label{fig:edge-reconfiguration}
	\vspace{-3mm}
\end{figure}

\begin{figure*}
	\begin{subfigure}[c]{0.24\linewidth}
		\includegraphics[width=\linewidth]{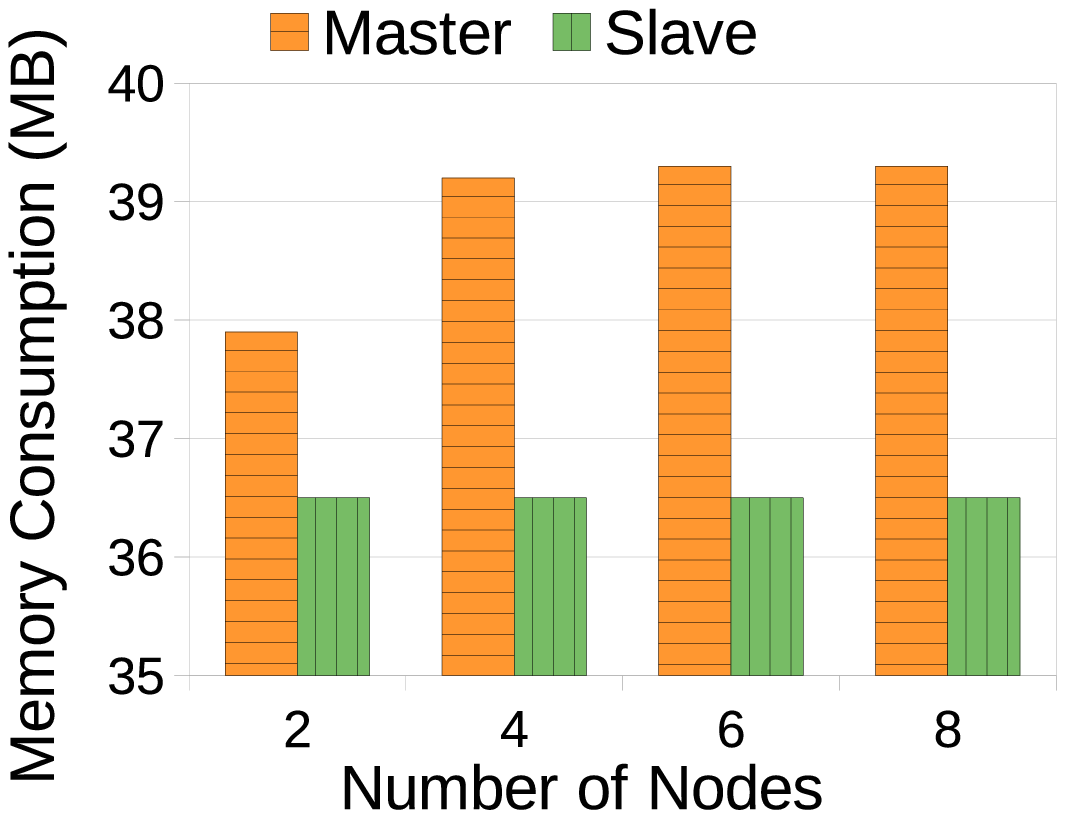}
		\caption{Avg. memory consumption}
		\label{fig:memory-consumption}
	\end{subfigure}
	\begin{subfigure}[c]{0.24\linewidth}
		\includegraphics[width=\linewidth]{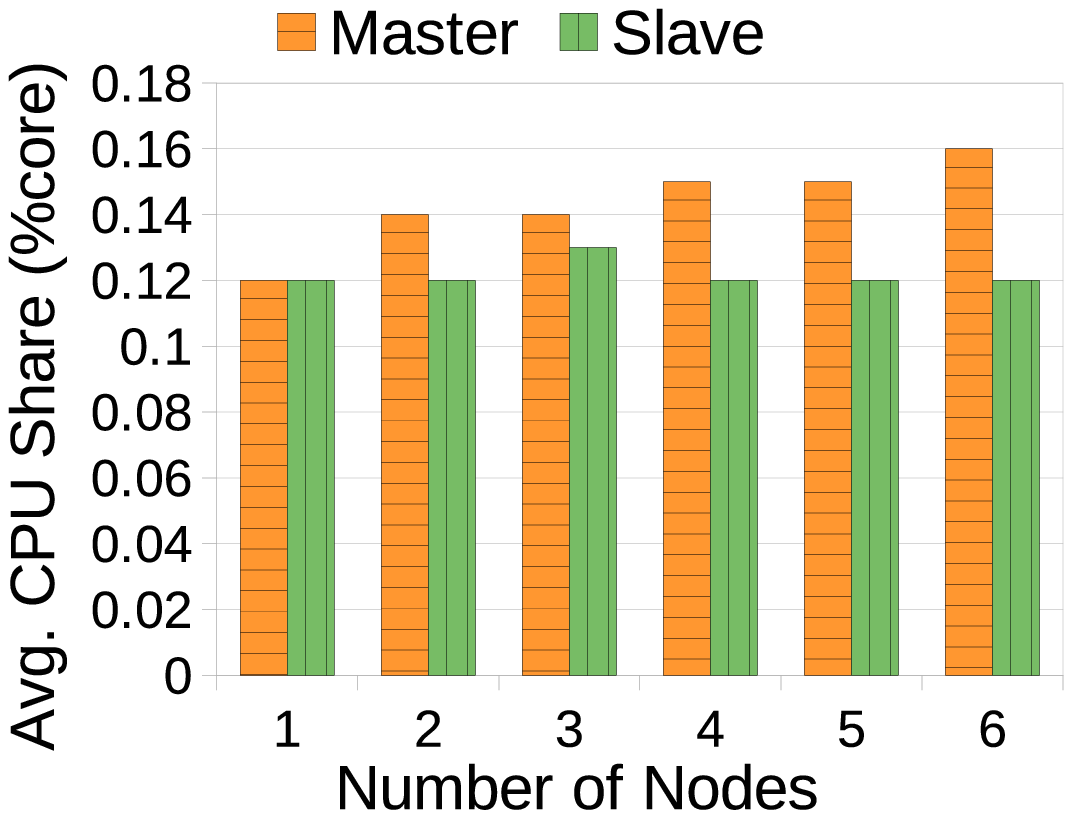}
		\caption{Avg. processor consumption}
		\label{fig:processy-consumption}
	\end{subfigure}
	\begin{subfigure}[c]{0.24\linewidth}
		\includegraphics[width=\linewidth]{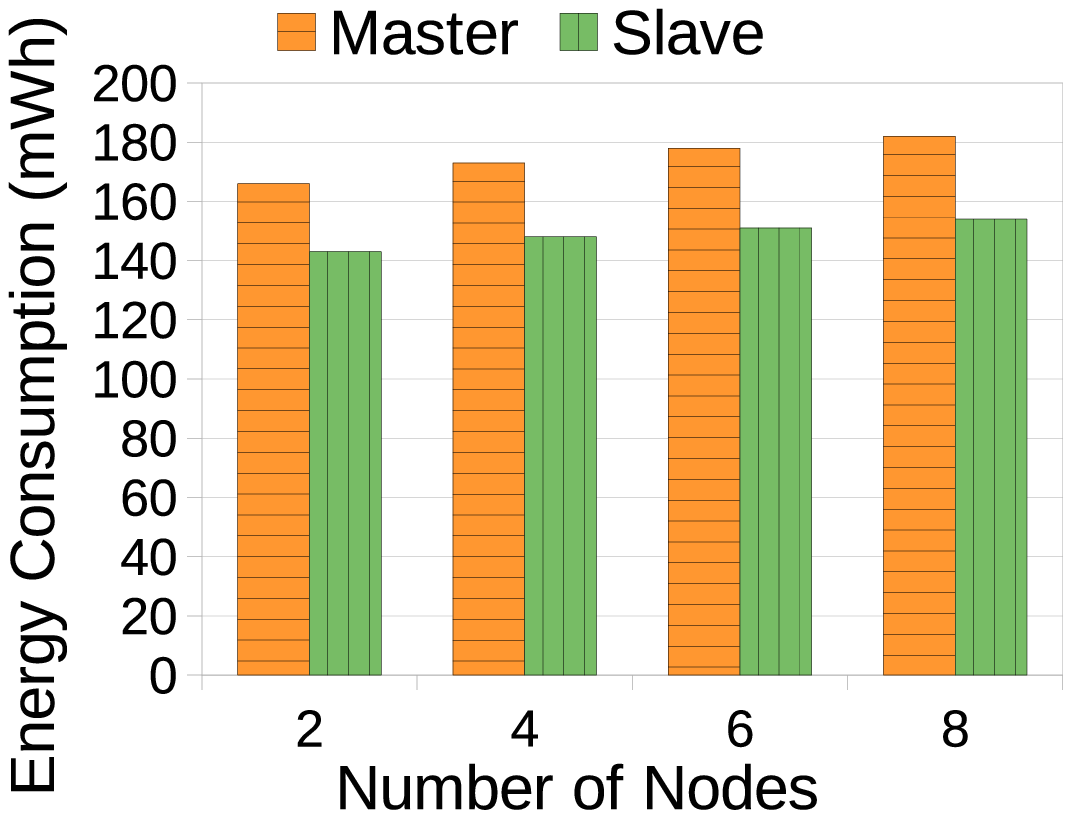}
		\caption{Avg. energy consumption}
		\label{fig:energy-consumption}
	\end{subfigure}
	\begin{subfigure}[c]{0.24\linewidth}
		\includegraphics[width=\linewidth]{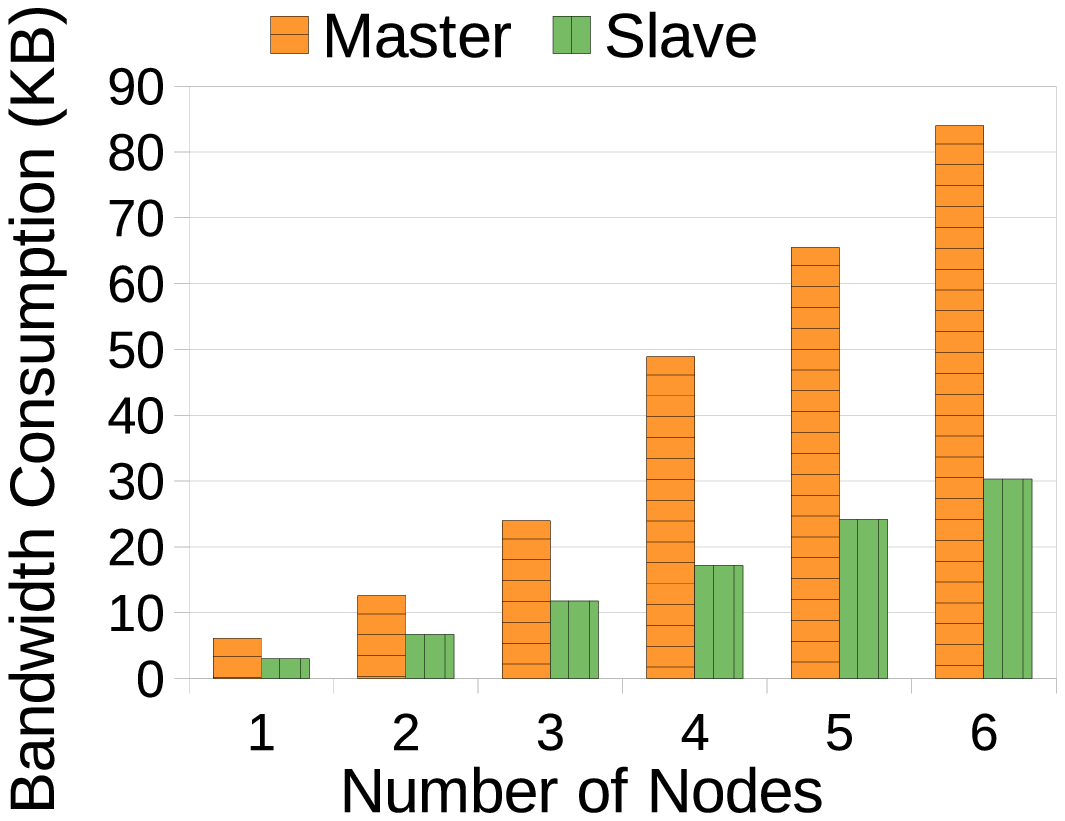}
		\caption{Avg. network traffic}
		\label{fig:network-consumption}
	\end{subfigure}
	\vspace{-10pt}
	\caption{Resources consumed by \ek{} on an Essential mobile phone.}
	\label{fig:resource-cosumption}
	\vspace{-3mm}
\end{figure*}

\subsection{\ek{} Resilience}
As the second step of evaluating the \ek{} internals, we investigate the resiliency of \ek{} in terms of replica departure (described in Section~\ref{sec:resiliency}). The results are depicted in Figure~\ref{fig:edge-reconfiguration}. In these experiments, we removed some replica nodes from the network to check how it affects the remaining cluster. In the case of 1-replica, the master serves as the sole replica. As soon as the master becomes unavailable, all the \ek{} slaves stop serving clients.  In multi replica scenarios, if a replica becomes unavailable, the rest of the \ek{} processes keep serving the clients as long as more than half of the configured replicas remain active. The replica departure is detected using the topology discovery. When a master detects a link failure with a replica, it waits for $4\times topology\ ping\ interval$ to select a new node to be a replica. 
In all the cases, we can see that the reconfiguration is completed within 60s. 

%
%

\subsection{Lightweight Mobile App Performance}
As the last step of evaluating the \ek{} internals, we use the Android Studio profiler to measure resource consumption of running \ek{} on mobile devices. Figure~\ref{fig:resource-cosumption} plots the memory consumption, processor consumption, energy consumption, network usage of \ek{} applications. The plots show that when running in master mode, an application consumes significantly higher memory and network traffic than running in slave mode. However, the processor usage and energy consumption are similar for master and slave modes. In slave mode, the memory consumption remains intact even when the number of nodes in the cluster increases. However, the network traffic increases with the number of devices in the cluster. In all cases, the memory consumption is below 40MB, which is negligible compared to 4GB internal memory of a mobile phone. Processor usage is within 0.16\%. \ek{} also consumes very low energy (0.17 watt per hour) compared to the battery capacity of 3050mAh. The results clearly indicate that \ek{} is very lightweight and suitable for running on mobile phones.

%
%
%
%
%
%
%
%

\subsection{Performance of Service Discovery}
\begin{figure}
	\centering
	\includegraphics[width=0.7\linewidth]{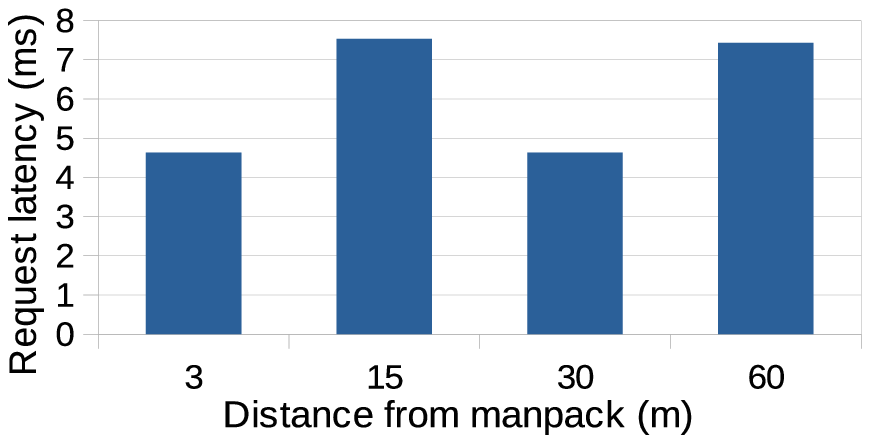}
	\caption{Service discovery API latency for 1-replica \ek{} cluster. The phones are connected to the NIST deployable manpack through LTE.}
	\label{fig:gypsum-latency}
	\vspace{-20pt}
\end{figure}

\begin{figure*}
	\begin{subfigure}[c]{0.32\linewidth}
		\includegraphics[width=\linewidth]{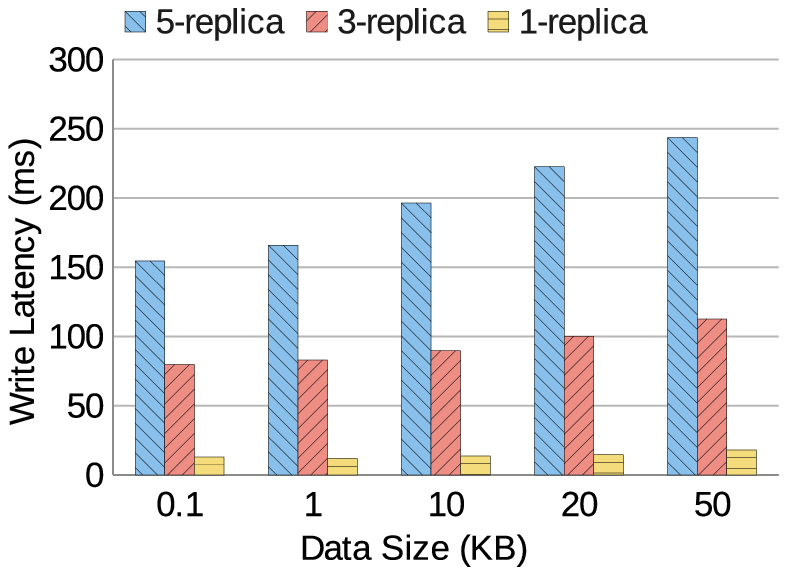}
		\caption{Request made on a replica}
		\label{fig:latency-write-master}
	\end{subfigure}
	\begin{subfigure}[c]{0.32\linewidth}
		\includegraphics[width=\linewidth]{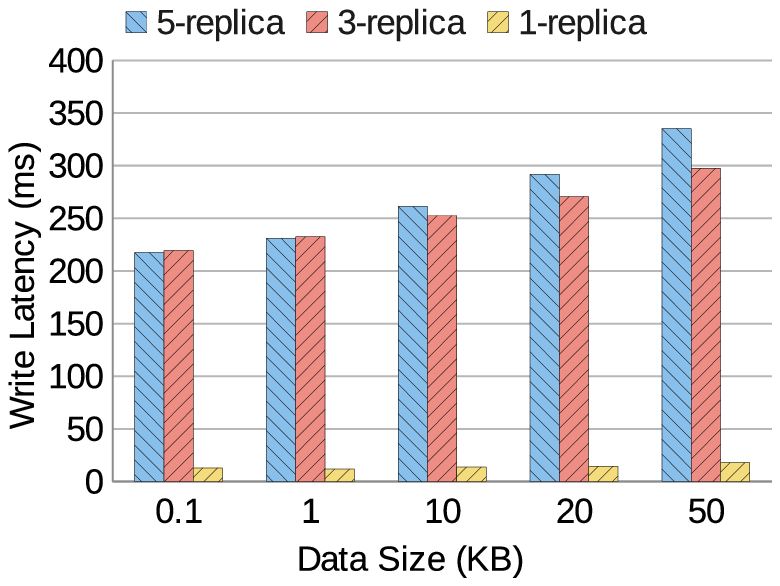}
		\caption{Request made on a slave}
		\label{fig:latency-write-client}
	\end{subfigure}
	\begin{subfigure}[c]{0.32\linewidth}
		\includegraphics[width=\linewidth]{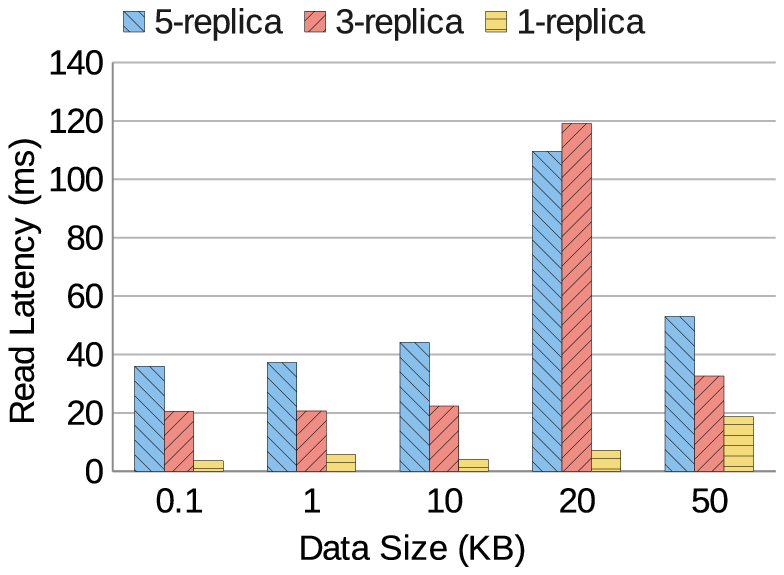}
		\caption{Request made on a replica}
		\label{fig:latency-read-master}
	\end{subfigure}
	\vspace{-10pt}
	\caption{Latency of client requests. We conducted experiments for multiple requests through MDFS application and obtained the average latency. In these experiments all the nodes are connected over WiFi.}
	\label{fig:request-latency}
	\vspace{-5mm}
\end{figure*}

After analyzing the \ek{} internals, we now assess the service discovery API provided to the client applications as described in Section~\ref{sec:service-discovery}. We conducted several experiments at the Gypsum open field (Figure~\ref{fig:gypsum_areal}) using LTE networks. The experiments are repeated to investigate the performance change with distance. We used MStorm as the client application to invoke service discovery API calls. The results are plotted in Figure~\ref{fig:gypsum-latency}. As shown, the delays are within the 8ms. Note that these requests are made to the \ek{} running at the local device, and most of these requests can be satisfied by looking at the local cache. Thus, the delay variance is caused because of the processing load on that device at this time. 

\subsection{Performance of MetaData Storage}
As the next step to evaluate the client API performance, we investigate the metadata storage performance as described in Section~\ref{sec:metadata-storage}. First, we investigate the latency incurred by metadata storage API calls made by MDFS. In a laboratory setup, all the nodes are connected over WiFi, where The manpack acts as the master. We vary the number of replicas to realize the effect of replica number on the request latency. Figure~\ref{fig:request-latency} plots the results. We repeated the experiments several times and took the average. Figure~\ref{fig:latency-write-master} and Figure~\ref{fig:latency-write-client} plot the delay incurred by \emph{putMetadata} service calls on a replica and slave device respectively. Figure~\ref{fig:latency-read-master} plots the average latency incurred by \emph{getMetadata} calls on a replica node. We can clearly observe that the latency is significantly higher when the request is made on a slave node as it involves network traffic. The plots also show that the read latency is lower than write latency as writing data requires consensus among the replicas. It also shows that the latency rises with an increase in the number of replicas as with more replicas, the consensus protocol mandates a higher number of message exchanges among the replicas.   

\begin{figure}[t]
	\centering
	\begin{subfigure}[c]{0.49\linewidth}
		\includegraphics[width=\linewidth]{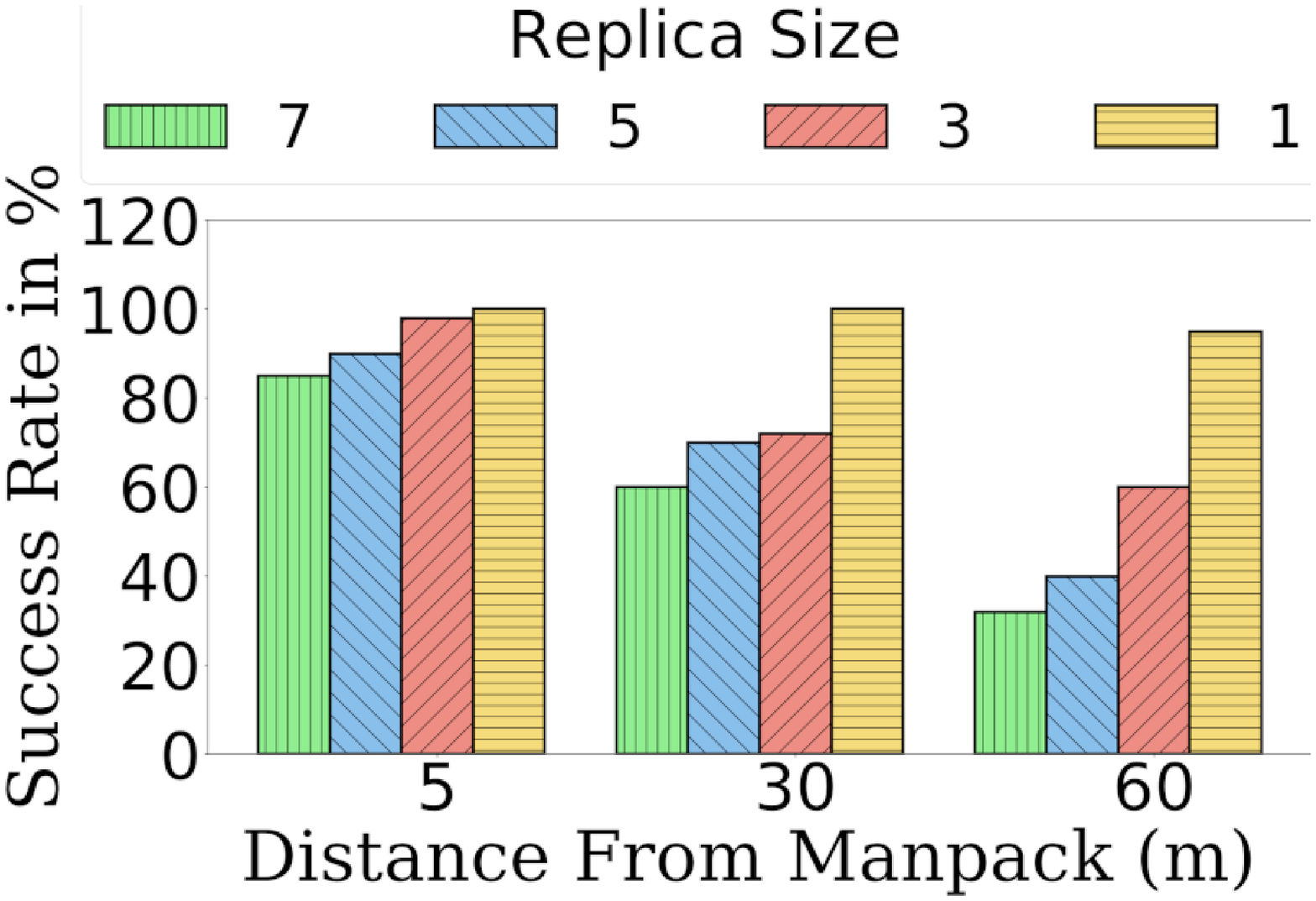}
		\caption{Avg. Success rate (\%)}
		\label{fig:res_mdfs_success}
	\end{subfigure}
	\begin{subfigure}[c]{0.49\linewidth}		
	    \includegraphics[width=\linewidth]{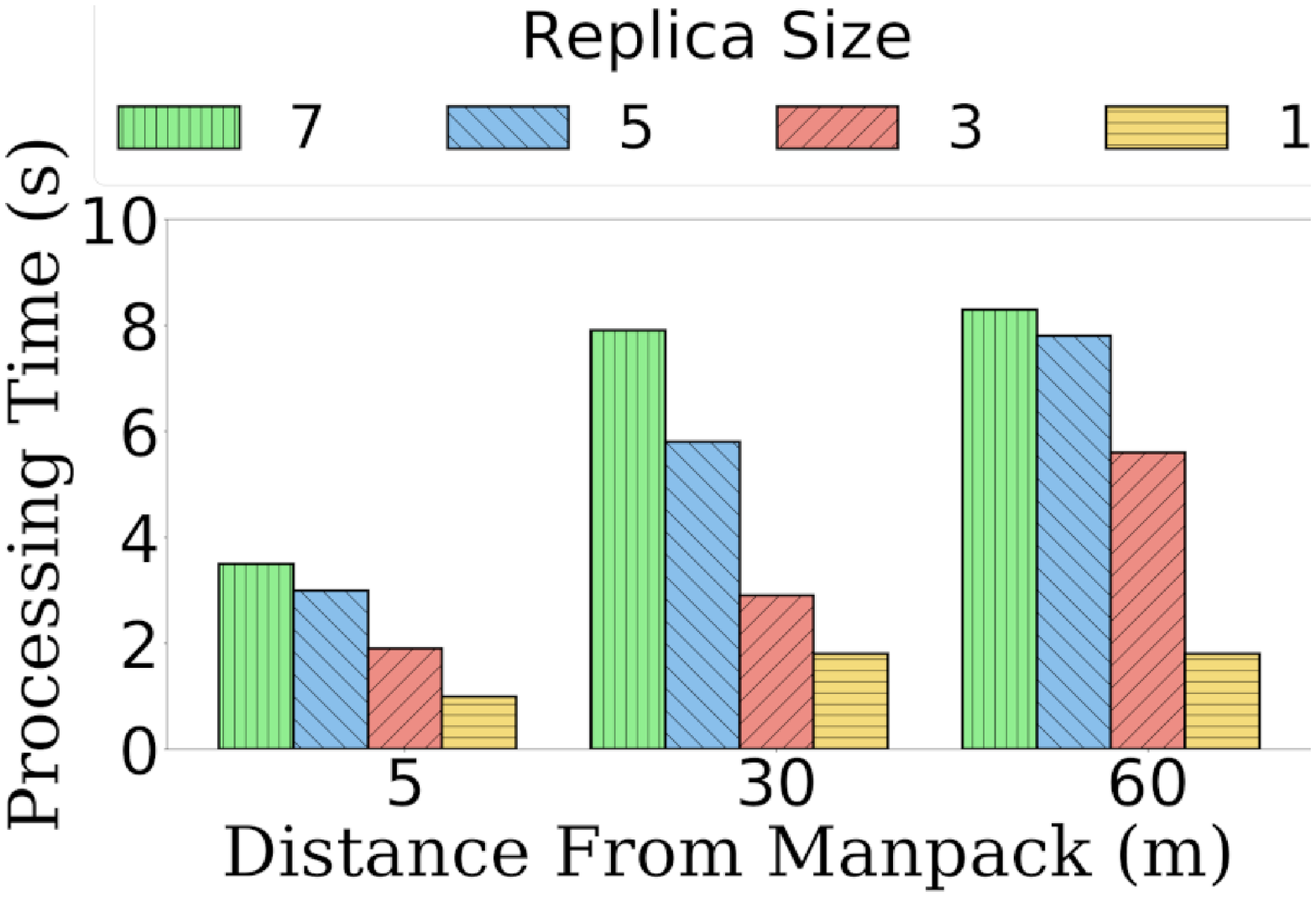}
		\caption{Avg. Processing time (s)}
		\label{fig:res_mdfs_time}
	\end{subfigure}
	\vspace{-10pt}
	\caption{Experiment results for MDFS file creation operation. Note that, the processing time includes the MDFS fragmentation delay.}
\end{figure}


We conducted another set of experiments to investigate the effect of link quality on the replica consensus. These experiments are conducted during Disaster City real-life deployment (Figure~\ref{fig:disaster_city_deployment}) using NIST manpack and mobile phones. We placed the mobile phones far from the manpack to vary the link qualities. We conducted the experiments in three varying network conditions: good link qualities (5m distance), moderate link qualities (30m distance), and poor link qualities (60m distance). We use MDFS client to invoke several \emph{putMetadata} API calls. We measure two metrics: success rate (Figure~\ref{fig:res_mdfs_success}) and total latency (Figure~\ref{fig:res_mdfs_time}). The success rate is the percentage of API calls successfully executed. The total latency is the delay incurred by 100 API calls.  We can see that the average response time increases significantly with the distance between manpack and phones from the plots. Note that beyond 100 m distance, the phones could not maintain good communication with the manpack, which resulted in unsuccessful consensus. It is always desired to go for higher replicas to achieve better resilience. However, we can observe that when the link qualities are poor, it is better to use fewer replicas as maintaining consensus for storing metadata over wireless links is very costly in terms of latency.

\section{Conclusions}\label{sec:conclusion}
In this paper, we present \ek, a resilient application coordination service for mobile edge networks. We designed \ek{} to provide device naming, application discovery, service coordination, metadata storage, and provide an edge status to the client applications. It provides cloud-like distributed coordination at the edge where mobile nodes can get the required coordination services when disconnected from the Internet. We have implemented EdgeKeeper on both Android and Linux platforms and tested it on real-life experiments on mobile edge networks formed by a group of first responders deployed after a disaster scenario. The tests showed that \ek{} is able to automatically discover devices in the network and form the edge network quickly. In the case of node failure and node departure, \ek{} is able to reconfigure the network and continue the operation of upper-layer applications. The tests also prove that \ek{} is lightweight and suitable to run on battery constrained mobile hand-held devices.

\bibliography{references}
\bibliographystyle{unsrt}

\end{document}